\documentclass[runningheads]{llncs}

\usepackage{url}
\usepackage{paralist}
\usepackage{fancyvrb}
\usepackage{epsfig}
\usepackage{epstopdf}
\usepackage[nomove]{cite}
\usepackage{booktabs} 
\usepackage{makecell}

\newcommand{\q}[1]{\lq\lq{}{}#1\rq\rq{}{}}

\begin{document}

\title{TweetsKB: A Public and Large-Scale RDF Corpus of Annotated Tweets}

\author{
    Pavlos Fafalios \and
    Vasileios Iosifidis \and
    Eirini Ntoutsi \and
    Stefan Dietze }
\institute{
    L3S Research Center, University of Hannover, Germany\\
    \email{\{fafalios, iosifidis, ntoutsi, dietze\}@L3S.de}}

\maketitle

\begin{abstract}
Publicly available social media archives facilitate research in a variety of fields, such as data science, sociology or the digital humanities, where Twitter has emerged as one of the most prominent sources. However, obtaining, archiving and annotating large amounts of tweets is costly. In this paper, we describe {\em TweetsKB}, a publicly available corpus of currently more than 1.5 billion tweets, spanning almost 5 years (Jan'13-Nov'17). Metadata information about the tweets as well as extracted entities, hashtags, user mentions and sentiment information are exposed using established RDF/S vocabularies. Next to a description of the extraction and annotation process, we present use cases to illustrate scenarios for entity-centric information exploration, data integration and knowledge discovery facilitated by {\em TweetsKB}.

\end{abstract}


\section{Introduction}
\label{sec:intro}

Social microblogging services have emerged as a primary forum to discuss and comment on breaking news and events happening around the world. Such user-generated content can be seen as a comprehensive documentation of the society and is of immense historical value for future generations \cite{bruns2016twitter}.

In particular, Twitter has been recognized as an important data source facilitating research in a variety of fields, such as data science, sociology, psychology or historical studies where researchers aim at understanding behavior, trends and opinions. While research usually focuses on particular topics or entities, such as persons, organizations, or products, entity-centric access and exploration methods are crucial \cite{weikum2011longitudinal}.

However, despite initiatives aiming at collecting and preserving such user-generated content (e.g., the Twitter Archive at the Library of Congress \cite{zimmer2015twitter}), the absence of publicly accessible archives which enable entity-centric exploration remains a major obstacle for research and reuse \cite{bruns2016twitter}, in particular for non-technical research disciplines lacking the skills and infrastructure for large-scale data harvesting and processing.

In this paper, we present {\em TweetsKB}, a public corpus of RDF data for a large collection of anonymized tweets. {\em TweetsKB} is unprecedented as it currently contains data for more than 1.5 billion tweets spanning almost 5 years, includes entity and sentiment annotations, and is exposed using established vocabularies in order to facilitate a variety of multi-aspect data exploration scenarios.

By providing a well-structured large-scale Twitter corpus using established W3C standards, we relieve data consumers from the computationally intensive process of extracting and processing tweets, and facilitate a number of data consumption and analytics scenarios including:
i) time-aware and entity-centric exploration of the Twitter archive \cite{fafalios2017jcdl},
ii) data integration by directly exploiting existing knowledge bases (like DBpedia) \cite{fafalios2017jcdl}, 
iii) multi-aspect entity-centric analysis and knowledge discovery w.r.t. features like entity popularity, attitude or relation with other entities \cite{fafalios2017tpdl}. 
In addition, the dataset can foster further research, for instance, in entity recommendation, event detection, topic evolution, and concept drift.

Next to describing the annotation process (entities, sentiments) and the access details (Section \ref{sec:corpus}), we present the applied schema (Section \ref{sec:model}) as well as use case scenarios and update and maintenance procedures (Section \ref{sec:usecases}). Finally, we discuss related works (Section \ref{sec:rw}) and conclude the paper (Section \ref{sec:concl}).


\section{Generating TweetsKB}
\label{sec:corpus}

{\em TweetsKB} is generated through the following steps:
i) tweet archival, filtering and processing,
ii) entity linking and sentiment extraction, and
iii) data lifting.
This section summarizes the above steps while the corresponding schema for step (iii) is described in the next section.

\subsection{Twitter Archival, Filtering and Processing}
The archive is facilitated by continuously harvesting tweets through the public Twitter streaming API since January 2013, accumulating more than 6 billion tweets up to now (December 2017).

\begin{figure*}[t]
    \centering
    \includegraphics[width=4.8in]{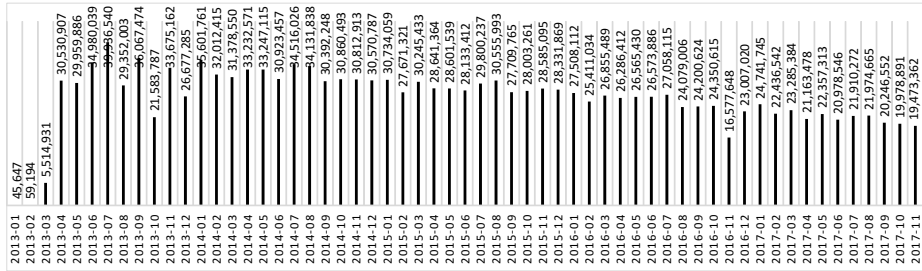}
    \vspace{-5mm}
    \caption{Number of tweets per month of the {\em TweetsKB} dataset.}
    \label{fig:tweetsPerMonth}
\end{figure*}

As part of the filtering step, we eliminate re-tweets and non-English tweets, which has reduced the number of tweets to about 1.8 billion tweets. In addition, we remove spam through a Multinomial Naive Bayes (MNB) classifier, trained on the HSpam dataset which has 94\% precision on spam labels \cite{sedhai2015hspam14}.
This removed about 10\% of the tweets, resulting in a final corpus of 1,560,096,518 tweets.
Figure \ref{fig:tweetsPerMonth} shows the number of tweets per month of the final dataset.

For each tweet, we exploit the following metadata: tweet id, post date, user who posted the tweet (username), favourite and retweet count (at the time of fetching the tweet\footnote{By exploiting the tweet IDs, one can retrieve the latest favourite and retweet counts (however, only in case the corresponding tweets have not been deleted and are still publicly accessible).}). 
We also extract hashtags (words starting with \#) and user mentions (words starting with @). 
For the sake of privacy, we anonymize the usernames and we do not provide the text of the tweets (nevertheless, one can still apply user-based aggregation and analysis tasks). However, actual tweet content and further information can be fetched through the tweet IDs.

\subsection{Entity Linking and Sentiment Extraction}
For the {\em entity linking} task, we used Yahoo's FEL tool \cite{blanco2015fast}. FEL is very fast and lightweight, and has been specially designed for linking entities from short texts to Wikipedia/DBpedia.
We set a confidence threshold of -3 which has been shown empirically to provide annotations of good quality, while we also store the confidence score of each extracted entity. Depending on the specific requirements with respect to precision and recall, data consumers can select suitable confidence ranges to consider when querying the data.

\begin{figure}[t]
    \centering
    \includegraphics[width=4.8in]{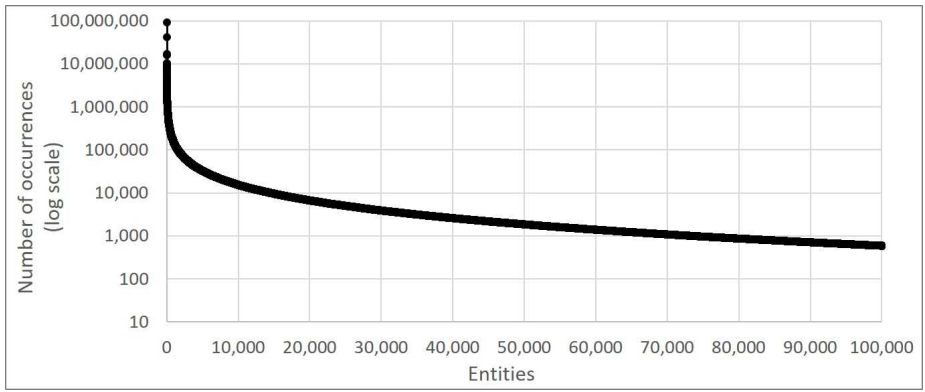}
    \vspace{-4mm}
    \caption{Distribution of top-100,000 entities.}
    \label{fig:entDistr}
\end{figure}

In total, about 1.4 million distinct entities were extracted from the entire corpus, while the average number of entities per tweet is about 1.3. 
Figure \ref{fig:entDistr} shows the distribution of the top-100,000 entities. 
There are around 15,000 entities with more than 10,000 occurrences, while there is a long tail of entities with less than 1,000 occurrences. 
Regarding their type, Table \ref{entityTypes} shows the distribution of the top-100,000 entities in some popular DBpedia types (the sets are not disjoint). We notice that around 20\% of the entities is of type {\em Person} and 15\% of type {\em Organization}. 

\begin{table}[]
\vspace{-1mm}
\centering
\caption{Overview of popular entity types of the top-100,000 entities.}
\label{entityTypes}
\begin{tabular}{p{6.5cm}r}
    \toprule
    DBpedia type                                & Number of distinct entities \\
    \midrule
    http://dbpedia.org/ontology/Person	        &   21,139 (21.1\%) \\
    http://dbpedia.org/ontology/Organisation    &	14,815 (14.8\%) \\
    http://dbpedia.org/ontology/Location	    &   8,215 (8,2\%)   \\
    http://dbpedia.org/ontology/Athlete	        &   5,192 (5.2\%)   \\
    http://dbpedia.org/ontology/Artist	        &   3,737 (3.7\%)   \\
    http://dbpedia.org/ontology/City	        &   2,563 (2.6\%)   \\
    http://dbpedia.org/ontology/Event	        &   510 (0.5\%)     \\
    http://dbpedia.org/ontology/Politician	    &   208 (0.2\%)     \\
  \bottomrule
\end{tabular}
\end{table}

For {\em sentiment analysis}, we used SentiStrength, a robust tool for sentiment strength detection on social web data \cite{thelwall2012sentiment}. SentiStrength assigns both a positive and a negative score to a short text, to account for both types of sentiments expressed at the same time. The value of a positive sentiment ranges from +1 for no positive to +5 for extremely positive. Similarly, negative sentiment ranges from -1 (no negative) to -5 (extremely negative). We normalized both scores in the range $[0, 1]$ using the formula: $score = (|sentimentValue| - 1) / 4)$. About 788 million tweets (50\%) have no sentiment ($score=0$ for both positive and negative sentiment). 

\subsubsection{Quality of Annotations} 
We evaluated the quality of the {\em entity annotations} produced by FEL using the ground truth dataset provided by the 2016 NEEL challenge of the 6th workshop on \q{Making Sense of Microposts} (\#Microposts2016)\footnote{\url{http://microposts2016.seas.upenn.edu/}} \cite{rizzo2016making}. The dataset consists of 9,289 English tweets of 2011, 2013, 2014, and 2015. We considered all tweets from the provided training, dev and test files, without applying any training on FEL.
The results are the following: {\em Precision} = 86\%, {\em Recall} = 39\%, {\em F1} = 54\%.
We notice that FEL achieves high precision, however recall is low. The reason is that FEL did not manage to recognize several difficult cases, like entities within hashtags and nicknames, which are common in Twitter due to the small number of allowed characters per tweet.
Nevertheless, FEL's performance is comparable to existing approaches \cite{rizzo2015making,rizzo2016making}.

Regarding {\em sentiment analysis}, we evaluated the accuracy of SentiStrength on tweets using two ground truth datasets: SemEval2017\footnote{\url{http://alt.qcri.org/semeval2017/task4/}} (Task 4, Subtask A) \cite{rosenthal2017semeval}, and TSentiment15\footnote{\url{https://l3s.de/~iosifidis/TSentiment15/}} \cite{iosifidis2017large}. 
The SemEval2017 dataset consists of 61,853 English tweets of 2013-2017 labeled as positive, negative, or neutral. We run the evaluation on all the provided training files (of 2013-2016) and the 2017 test file.
SentiStrength achieved the following scores:  {\em  AvgRec = 0.54} (recall averaged across the positive, negative, and neutral classes \cite{sebastiani2015axiomatically}), {\em $F1^{PN}$ = 0.52} (F1 averaged across the positive and negative classes), {\em Accuracy} = 0.57. 
The performance of SentiStrength is good considering that this is a multi-class classification problem. Moreover, the user can achieve higher precision by selecting only tweets with high positive or negative SentiStrength score.
Regarding TSentiment15, this dataset contains 2,527,753 English tweets of 2015 labeled only with positive and negative classes (exploiting emoticons and a sentiment lexicon \cite{iosifidis2017large}).
 SentiStrength achieved the following scores: {\em $F1^{PN}$} = 0.80, {\em Accuracy} = 0.91. 
Here we notice that SentiStrength achieves very good performance.

\subsection{Data Lifting \& Availability}
We generated RDF triples in the N3 format applying the RDF/S model described in the next section. The total number of triples is more than 48 billion. Table \ref{stats} summarizes the key statistics of the generated dataset.  The source code used for triplifying the data is available as open source on GitHub\footnote{\label{fn:git}\url{https://github.com/iosifidisvasileios/AnnotatedTweets2RDF}}.

\begin{table}[]
\centering
\caption{Key statistics of {\em TweeetsKB}.}
\label{stats}
\begin{tabular}{p{5.5cm}r}
    \toprule
    Number of tweets:                   & 1,560,096,518 \\ 
    Number of distinct users:           & 125,104,569   \\ 
    Number of distinct hashtags:        & 40,815,854   \\ 
    Number of distinct user mentions:   & 81,238,852   \\ 
    Number of distinct entities:        & 1,428,236   \\ 
    Number of tweets with sentiment:    & 772,044,599  \\ 
    Number of RDF triples:              & 48,207,277,042   \\ 
  \bottomrule
\end{tabular}
\end{table}

{\em TweetsKB} is available as N3 files (split by month) through the Zenodo data repository (DOI: 10.5281/zenodo.573852)\footnote{\url{https://zenodo.org/record/573852}}, under a {\em Creative Commons Attribution 4.0} license. The dataset has been also registered at {\tt datahub.ckan.io}\footnote{\url{https://datahub.ckan.io/dataset/tweetskb}}.
Sample files, example queries and more information are available through {\em TweetsKB}'s home page\footnote{\url{http://l3s.de/tweetsKB/}}.
For demonstration purposes, we have also set up a public SPARQL endpoint, currently containing a subset of about 5\% of the dataset\footnote{\url{http://l3s.de/tweetsKB/endpoint/} (Graph IRI: \url{http://l3s.de/tweetsKB/})}.

\subsection{Runtime for Annotation and Triplification}
The time for annotating the tweets and generating the RDF triples depends on several factors including the dataset volume, the used computing infrastructure as well as the available resources and the load of the cluster during the analysis time. The Hadoop cluster used for creating {\em TweetsKB} consists of 40 computer nodes with a total of 504 CPU cores and 6,784 GB RAM. The most time consuming task is entity linking where we annotated on average 4.8M tweets per minute using FEL, while SentiStrength annotated almost 6M tweets per minute. Finally, for the generation of the RDF triples we processed 14M tweets per minute on average.


\section{RDF/S Model for Annotated Tweets}
\label{sec:model}

\begin{figure*}[t]
    \centering
    \fbox{\includegraphics[width=4.7in]{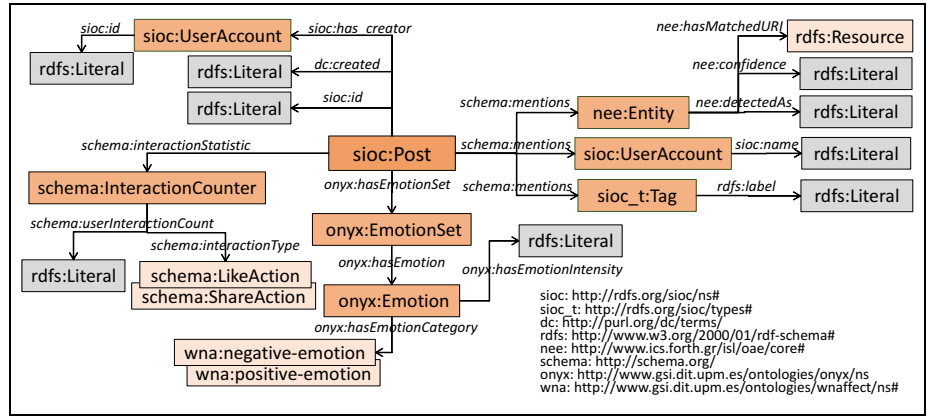}}
    \vspace{-4mm}
    \caption{An RDF/S model for describing metadata and annotation information for a collection of tweets.}
    \label{fig:model}
\end{figure*}

Our schema, depicted in Figure \ref{fig:model}, exploits terms from established vocabularies, most notably SIOC (Semantically-Interlinked Online Communities) core ontology \cite{breslin2006sioc} and schema.org \cite{ronallo2012html5}. 
The selection of the vocabularies was based on the following objectives: i) avoiding schema violations, ii) enabling data interoperability through term reuse, iii) having dereferenceable URIs, iv) extensibility. 
Next to modeling data in our corpus, the proposed schema can be applied over any annotated social media archive (not only tweets), and can be easily extended for describing additional information related to archived social media data and extracted annotations.

A tweet is associated with six main types of elements:
(1) {\em general tweet metadata},
(2) {\em entity mentions},
(3) {\em user mentions},
(4) {\em hashtag mentions},
(5) {\em sentiment scores},
(6) {\em interaction statistics} (values expressing how users have interacted with the tweet, like favourite and retweet count).
We use the property \textsf{schema:mentions} from schema.org\footnote{\url{http://schema.org/}} for associating a tweet with a mentioned entity, user or hashtag. We exploit schema.org due to its wide acceptance and less strict domain/range bindings which facilitate reuse and combination with other schemas, by avoiding schema violations.

For general {\em metadata}, we exploit SIOC as an established vocabulary for representing social Web data\footnote{Specification available at: \url{http://rdfs.org/sioc/spec/}}. The class \textsf{sioc:Post} represents a tweet, while \textsf{sioc:UserAccount} a Twitter user.

An {\em entity mention} is represented through the Open NEE (Named Entity Extraction) model \cite{fafalios2015ijait} which is an extension of the Open Annotation data model \cite{sanderson2013open} and enables the representation of entity annotation results. For each recognized entity, we store its surface form, URI and confidence score. A {\em user mention} simply refers to a particular \textsf{sioc:UserAccount}, while for {\em hashtag mentions} we use the class \textsf{sioc\_t:Tag} of the SIOC Types Ontology Module\footnote{\url{http://rdfs.org/sioc/types\#}}.

For expressing {\em sentiments}, we use the Onyx ontology\footnote{\url{https://www.gsi.dit.upm.es/ontologies/onyx/}} \cite{sanchez2016onyx}. Through the class \textsf{onyx:EmotionSet} we associate a tweet with a set of emotions (\textsf{onyx:Emotion}). Note that the original domain of property \textsf{onyx:has\-Emo\-tion\-Set} is \textsf{owl:Thing}, which is compatible with our use as property of \textsf{sioc:Post}. The property \textsf{onyx:\-has\-Emo\-tion\-Ca\-te\-go\-ry} defines the emotion type, which is either \textsf{negative-emotion} or \textsf{positive-emo\-tion} as defined by the WordNet-Affect Taxonomy\footnote{\url{http://www.gsi.dit.upm.es/ontologies/wnaffect/}} and is quantified through \textsf{onyx:\-has\-Emo\-tion\-Inte\-nsi\-ty}.

Finally, for representing aggregated {\em interactions}, we use the class \textsf{Inte\-ra\-cti\-onCounter} of schema.org. We distinguish \textsf{schema:Like\-Action} (for the favourite count) or \textsf{schema:\-ShareAction} (for the retweet count) as valid interaction types. 

\begin{figure}[t]
    \centering
    \fbox{\includegraphics[width=4.7in]{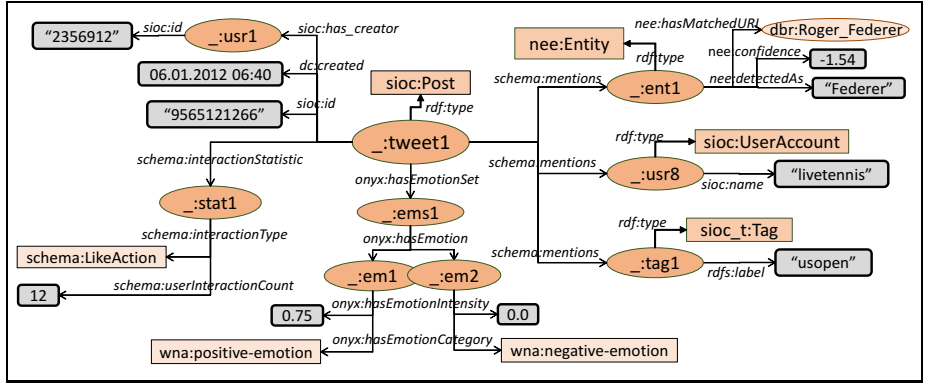}}
    \vspace{-4mm}
    \caption{Instantiation example of the RDF/S model.}
    \label{fig:modelExample}
\end{figure}

Figure \ref{fig:modelExample} depicts a set of instances for a single tweet. In this example, the tweet mentions one user account (@livetennis) and one hashtag (\#usopen), while the entity name {\em \q{Federer}} was detected, referring probably to the tennis player {\em Roger Federer}  (with confidence score $-1.54$). Moreover, we see that the tweet has a positive sentiment of 0.75, no negative sentiment, while it has been marked as \q{favourite} 12 times. 

\section{Use Cases and Sustainability}
\label{sec:usecases}

\subsection{Scenarios and Queries}

\renewcommand{\figurename}{Listing}
\setcounter{figure}{0}

Typical scenarios facilitated by {\em TweetsKB} include:

\vspace{1mm} \noindent
{\bf Advanced Exploration and Data Integration.}
By exploiting tweet metadata, extracted entities, sentiment values, and temporal information, 
one can run sophisticated queries that can also directly (at query-execution time) integrate information from external knowledge bases like DBpedia.
For example, Listing \ref{fig:advExpl1} shows a SPARQL query obtaining popular tweets in 2016 (with more than 100 retweets) mentioning {\em German politicians} with strong negative sentiment ($\geq 0.75$).
The query exploits extracted entities, sentiments, and interaction statistics, while it uses query federation to access DBpedia for retrieving the list of German politicians and their birth place.

\begin{figure}[th]
\vspace{-3mm}
\centering
\scriptsize
\begin{Verbatim}[frame=lines,numbers=left,numbersep=1pt]
SELECT DISTINCT ?tweetID ?sentNegScore ?retweetCount ?politician ?birthPlace WHERE {
  SERVICE <http://dbpedia.org/sparql> {
    ?politician dc:subject dbc:German_politicians ; dbo:birthPlace ?birthPlace }
  ?tweet a sioc:Post ; dc:created ?date ; sioc:id ?tweetID FILTER(year(?date) = 2016) .
  ?tweet schema:mentions ?entity . ?entity a nee:Entity ; nee:hasMatchedURI ?politician .
  ?tweet schema:interactionStatistic ?stat . ?stat schema:interactionType schema:ShareAction .
  ?stat schema:userInteractionCount ?retweetCount FILTER(?retweetCount > 100) .
  ?tweet onyx:hasEmotionSet ?emotSet . ?emotSet onyx:hasEmotion ?emot .
  ?emot onyx:hasEmotionCategory wna:negative-emotion ;
        onyx:hasEmotionIntensity ?sentNegScore FILTER (?sentNegScore >= 0.75) }
\end{Verbatim}
\vspace{-5mm}
\caption{SPARQL query for retrieving popular tweets in 2016 mentioning German politicians with strong negative sentiment.}
\label{fig:advExpl1}
\vspace{-3mm}
\end{figure}

Listing \ref{fig:advExpl2} shows a query that combines extracted entities with hashtags. 
The query requests the top-50 hashtags co-occurring with the entity {\em Refugee} (\url{http://dbpedia.org/resource/Refugee}) in tweets of 2016. 
The result contains, among others, the following hashtags: \#auspol, \#asylum, \#Nau\-ru, \#Gree\-ce, \#Let\-Them\-Stay, \#Bring\-Them\-Here.

\begin{figure}[th]
\vspace{-3mm}
\centering
\scriptsize
\begin{Verbatim}[frame=lines,numbers=left,numbersep=1pt]
SELECT ?hastagLabel (count(distinct ?tweet) as ?num) WHERE {
  ?tweet dc:created ?date FILTER(year(?date) = 2016) .
  ?tweet schema:mentions ?entity .
  ?entity a nee:Entity ; nee:hasMatchedURI dbr:Refugee .
  ?tweet schema:mentions ?hashtag.
  ?hashtag a sioc:Tag ; rdfs:label ?hastagLabel 
} GROUP BY ?hastagLabel ORDER BY DESC(?num) LIMIT 50
\end{Verbatim}
\vspace{-5mm}
\caption{SPARQL query for retrieving the top-50 hashtags co-occurring with the entity {\em Refugee} in tweets of 2016.}
\label{fig:advExpl2}
\vspace{-3mm}
\end{figure}

\vspace{1mm} \noindent
{\bf Temporal Entity Analytics.}
The work in \cite{fafalios2017tpdl} has proposed a set of measures that allow studying how entities are reflected in a social media archive and how entity-related information evolves over time.
Given an entity and a time period, the proposed measures capture the following entity aspects:
{\em popularity},
{\em attitude} (predominant sentiment),
{\em sentimentality} (magnitude of sentiment),
{\em controversiality}, and
{\em connectedness} to other entities (entity-to-entity connectedness and k-network).
Such time-series data can be easily computed by running SPARQL queries on {\em TweetsKB}.
For example, the query in Listing \ref{fig:obamaPopularity} retrieves the monthly popularity  of {\em Alexis Tsipras} (Greek prime minister) in Twitter in 2015 (using Formula 1 of \cite{fafalios2017tpdl}). The result of this query shows that the number of tweets increased significantly in June and July, likely to be caused by the Greek bailout referendum that was held in July 2015, following the bank holiday and capital controls of June 2015.

\begin{figure}[th]
\vspace{-3mm}
\centering
\scriptsize
\begin{Verbatim}[frame=lines,numbers=left,numbersep=1pt]
SELECT ?month xsd:double(?cEnt)/xsd:double(?cAll)
WHERE {
{ SELECT (month(?date) AS ?month) (count(?tweet) AS ?cAll) WHERE {
   ?tweet a sioc:Post ; dc:created ?date FILTER(year(?date) = 2015)
  } GROUP BY month(?date) }
{ SELECT (month(?date) AS ?month) (count(?tweet) AS ?cEnt) WHERE {
   ?tweet a sioc:Post ; dc:created ?date FILTER(year(?date) = 2015) .
   ?tweet schema:mentions ?entity . 
   ?entity a nee:Entity ; nee:hasMatchedURI dbr:Alexis_Tsipras
  } GROUP BY month(?date) }
} ORDER BY ?month
\end{Verbatim}
\vspace{-5mm}
\caption{SPARQL query for retrieving the monthly popularity of {\em Alexis Tsiprats} (Greek prime minister) in tweets in 2015 (using Formula 1 of \cite{fafalios2017tpdl}).}
\label{fig:obamaPopularity}
\vspace{-3mm}
\end{figure}

\vspace{1mm} \noindent
{\bf Time and Social Aware Entity Recommendations.}
Recent works have shown that entity recommendation is time-dependent,
while the co-occurrence of entities in documents of a given time period is
a strong indicator of their relatedness during that period and
thus should be taken into consideration \cite{zhang2016probabilistic,tran2017beyond}.
By querying {\em TweetsKB},
we can find entities of a specific type, or having some specific characteristics,
that co-occur frequently with a query entity in a specific time period,
a useful indicator for temporal prior probabilities when implementing time- and social-aware entity recommendations.
For example, the query in Listing \ref{fig:queryCooccured} retrieves the top-5 politicians
co-occurring with {\em Barack Obama} in tweets of summer 2016.
Here one could also follow a more sophisticated approach, e.g., by also considering
the inverse tweet frequency of the top co-occurred entities.

\begin{figure}[th]
\vspace{-3mm}
\centering \scriptsize
\begin{Verbatim}[frame=lines,numbers=left,numbersep=1pt]
SELECT ?politician (count(distinct ?tweet) as ?num) WHERE {
  SERVICE <http://dbpedia.org/sparql> {
     ?politician a dbo:Politician }
  ?tweet a sioc:Post ; dc:created ?date FILTER(?date >= "2016-06-01"^^xsd:date &&
                                               ?date <= "2016-08-30"^^xsd:date) .
  ?tweet schema:mentions ?entity .
  ?entity a nee:Entity ; nee:hasMatchedURI dbr:Barack_Obama .
  ?tweet schema:mentions ?entityPolit.
  ?entityPolit nee:hasMatchedURI ?politician FILTER (?politician != dbr:Barack_Obama)
} GROUP BY ?politician ORDER BY DESC(?num) LIMIT 5
\end{Verbatim}
\vspace{-5mm}
\caption{SPARQL query for retrieving the top-5 politicians
co-occurring with {\em Barack Obama} in tweets of summer 2016.}
\label{fig:queryCooccured}
\vspace{-3mm}
\end{figure}

\vspace{2mm} \noindent
{\bf Data Mining and Information Discovery}
Data mining techniques allow the extraction of useful and previously unknown information from raw data. By querying {\em TweetsKB} we can generate time series for a specific entity of interest modeling the temporal evolution of the entity w.r.t. different tracked dimensions like sentiment, popularity, or interactivity. Such multi-dimensional time-series can be used in a plethora of data mining tasks like entity forecasting (predicting entity-related features) \cite{saleiro2016learning}, network-analysis (find communities and influential entities) \cite{rossi2015spread}, stream mining (sentiment analysis over data streams) \cite{iosifidis2017sentiment,spiliopoulou2016opinion}, or change detection (e.g., detection of critical time-points) \cite{liu2013change}. 

Thus, research in a range of fields is facilitated through the public availability of well-annotated Twitter data.
Note also that the availability of publicly available datasets is a requirement for the data mining community and will allow not only the development of new methods but also for valid comparisons among existing methods, while existing repositories, e.g., UCI\footnote{\url{http://archive.ics.uci.edu/ml/}}, lack of big, volatile and complex data.

\subsection{Sustainability, Maintenance and Extensibility}

The dataset has seen adoption and facilitated research in inter-disciplinary research projects such as ALEXANDRIA\footnote{\url{http://alexandria-project.eu/}} and AFEL\footnote{\url{http://afel-project.eu/}}, involving researchers from a variety of organizations and research fields \cite{fafalios2017tpdl,kowald2017temporal,fafalios2017jcdl,tran15semanAnnot}.
With respect to ensuring long-term sustainability, we anticipate that reuse and establishing of a user community for the corpus is crucial. While the aforementioned activities have already facilitated access and reuse, the corpus will be further advertised through interdisciplinary networks and events (like the Web Science Trust\footnote{\url{http://www.webscience.org/}}). Besides, the use of Zenodo for depositing the dataset, as well as its registration at {\tt datahub.ckan.io}, makes it citable and web findable.

Maintenance of the corpus will be facilitated through the continuous process of crawling 1\% of all tweets (running since January 2013) through the public Twitter API and storing obtained data within the local Hadoop cluster at L3S Research Center.
The annotation and triplification process (Section \ref{sec:corpus}) will be periodically (quarterly) repeated in order to incrementally expand the corpus and ensure its currentness, one of the requirements for many of the envisaged use cases of the dataset. While this will permanently increase the population of the dataset, the schema itself is extensible and facilitates the enrichment of tweets with additional information, for instance, to add information about the users involved in particular interactions (retweets, likes) or additional information about involved entities or references/URLs.
Depending on the investigated research questions, it is anticipated that this kind of enrichment is essential, at least for parts of the corpus, i.e. for specific time periods or topics.

Next to the reuse of {\em TweetsKB}, we also publish the source code used for triplifying the data (see Footnote \ref{fn:git}), to enable third parties establishing and sharing similar corpora, for instance, focused Twitter crawls for certain topics.


\section{Related Work}
\label{sec:rw}

There is a plethora of works on modeling social media data as well as on semantic-based information access and mining semantics from social media streams (see \cite{bontcheva2014making} for a survey). 
There are also Twitter datasets provided by specific communities for research and experimentation in specific research problems, like the \q{Making Sense of Microposts} series of workshops \cite{rizzo2015making,rizzo2016making}, or the \q{Sentiment Analysis in Twitter} tasks of the International Workshop on Semantic Evaluation \cite{nakov2016semeval,rosenthal2017semeval}.
Below we discuss works that exploit Semantic Web technologies for representing and querying social media data.

Twarql \cite{mendes2010twarql} is an infrastructure which translates microblog posts from Twitter as Linked Data in real-time. Similar to our approach, Twarql extracts entity, hashtag and user mentions, and the extracted content is encoded in RDF. The authors tested their approach using a small collection of 511,147 tweets related to iPad\footnote{The dataset is not currently available (as of March 15, 2018).}.
SMOB \cite{passant2010overview} is a platform for distributed microblogging which combines Social Web principles and Semantic Web technologies. SMOB relies on ontologies for representing microblog posts, hubs for distributed exchanging information, and components for linking the posts with other resources.
TwitLogic \cite{shinavier2010real} is a semantic data aggregator which provides a set of syntax conventions for embedding various structured content in microblog posts. It also provides a schema for user-driven data and associated metadata which enables the translation of microblog streams into RDF streams. 
The work in \cite{sahito2011weaving} also discusses an approach to annotate and triplify tweets.
However, none of the above works provides a large-scale and publicly available RDF corpus of annotated tweets.


\section{Conclusion}
\label{sec:concl}

We have presented a large-scale Twitter archive which includes entity and sentiment annotations and is exposed using established vocabularies and standards. Data includes more than 48 billion triples, describing metadata and annotation information for more than 1.5 billion tweets spanning almost 5 years. Next to the corpus itself, the proposed schema facilitates further extension and the generation of similar, focused corpora, e.g. for specific geographic or temporal regions, or targeting selected topics. 

We believe that this dataset can foster further research in a plethora of research problems, like event detection, topic evolution, concept drift, and prediction of entity-related features, while it can facilitate research in other communities and disciplines, like sociology and digital humanities.

\subsection*{Acknowledgements}
The work was partially funded by the European Commission for the ERC Advanced Grant ALEXANDRIA under grant No. 339233 and the H2020 Grant No. 687916 (AFEL project), and by the German Research Foundation (DFG) project OSCAR (Opinion Stream Classification with Ensembles and Active leaRners).

\bibliographystyle{abbrv}
\bibliography{ESWC_TweetsKB__BIB}

\end{document}